\documentstyle[eqsecnum,aps]{revtex}

   \def\lesssim{\mathrel{\hbox{\rlap{\hbox{\lower4pt\hbox{$\sim$}}}\hbox{$<$}}}}
   \def\gtrsim{\mathrel{\hbox{\rlap{\hbox{\lower4pt\hbox{$\sim$}}}\hbox{$>$}}}}

   \newcommand{\bm}[1]{\mbox{\boldmath$#1$}}
   \newcommand{\mal}[1]{\stackrel{_\circ}{#1}}
   \newcommand{\kaco}[1]{\left\langle{#1}\right\rangle}
   \newcommand{\psm}{p^<}
   \newcommand{\pla}{p^>}
   \renewcommand{\thesection}{\arabic{section}}

\begin{document}
\title{How is the local-scale gravitational instability 
influenced by the surrounding large-scale structure formation?}

\author{Masahiro Takada\footnotemark[1]  and    
Toshifumi Futamase\footnotemark[2]}

\address{\small\sl Department of Astronomy, Faculty of Science,
Tohoku University, Sendai 980-8578, Japan}
\footnotetext[1]{Electronic address: takada@astr.tohoku.ac.jp}
\footnotetext[2]{Electronic address: tof@astr.tohoku.ac.jp}

\maketitle

\bigskip

%
%
%
%
%
%

\begin{abstract}

We develop the formalism 
to investigate the relation between the evolution of the large-scale
(quasi) linear structure and that of the small-scale nonlinear
structure in Newtonian cosmology within the Lagrangian framework. 
In doing so, we first derive the standard Friedmann  
expansion law using the averaging procedure over the present horizon 
scale. 
Then the large-scale (quasi) linear flow is defined by averaging the full 
trajectory field over a large-scale domain, but much smaller than the
horizon scale. 
The rest of the full trajectory field is supposed to describe 
small-scale nonlinear dynamics. We obtain the evolution equations for
the large-scale and small-scale parts of the trajectory field. 
These are coupled to each other in most general situations. 

It is shown that if the shear deformation of fluid elements 
is ignored in the averaged large-scale dynamics, the small-scale 
dynamics is described by Newtonian dynamics in an effective
 Friedmann-Robertson-Walker (FRW) background  with a local scale factor.
The local scale factor is defined by the  sum of the global scale factor 
and the expansion deformation of the averaged large-scale displacement field.   
This means that the evolution of small-scale fluctuations 
is influenced by the surrounding large-scale structure through 
the modification of FRW scale factor. 
The effect might play an important role in the structure formation
scenario.    
Furthermore, it is argued that the so-called {\it optimized} 
or {\it truncated} Lagrangian perturbation theory is a good
approximation in investigating the large-scale structure formation up to 
the  quasi 
nonlinear regime,  even when the small-scale fluctuations 
are in the non-linear regime. 

\end{abstract} 

\vspace{2em}

\section{Introduction}
The observation of the anisotropy of the cosmic microwave background (CMB) 
indicates that the universe is remarkably isotropic on the present horizon
scale. 
Thus it is natural to describe the horizon scale spatial geometry
of the universe by a homogeneous and isotropic metric, namely, the
Friedmann-Robertson-Walker (FRW) model. 
However, the real universe is neither isotropic nor homogeneous on  local
scales and has a hierarchical structure such as  galaxies,
clusters of galaxies, superclusters of galaxies and so on. 
It has been naively regarded that the FRW model is a large scale average 
of a locally inhomogeneous real universe.
There have been several studies in this direction in general relativity
\cite{futamase88,futamase89,futamase96,kasai95,russ,carfora}.

Aside from such a fundamental problem, there is an interesting and
practical problem associated with inhomogeneities across various scales. 
Can one ask if the formation of small-scale structures is influenced by 
the gravitational effect of structures with larger scales?
Such an environmental effect may be important and even essential 
to clarify the process of the hierarchical structure formation. 

This is the problem we attack in the present paper. Namely, we develop
the formalism to investigate the gravitational instability in general 
situations where the large-scale linear and the 
small-scale non-linear fluctuations coexist. 
If one uses the N-body simulation to answer the above question,  
one needs high spatial resolution over a very large box comparable with 
the horizon scale,  which may be well above the ability of the present
computer. 
However, it seems reasonable to regard the situation such as 
local nonlinear structures are superimposed on a smoothed large-scale 
linear structure and the large-scale dynamics may well be treated 
by Zel'dovich-type approximations for an usual power spectrum. 
This suggests us to adopt  an analytical approach based on the 
Lagrangian perturbation theory in Newtonian cosmology.
  
The reason why we consider Newtonian cosmology is partly 
because of its simplicity and partly because the Newtonian cosmology 
is a good approximation to a realistic inhomogeneous universe. In fact 
we have shown that the Newtonian cosmology 
in the relativistic framework is a good approximation even for the
perturbations not only inside but also beyond the present horizon scale 
\cite{tf,tfptp}.

The reason why we work within the Lagrangian framework is 
because it seems easy to introduce the averaging process. 
This is essential in our formalism because we are going to 
define the global expansion law as well as the large-scale smoothed
trajectory field defined by averaging. In fact Buchert and Ehlers studied 
the averaging problem in Newtonian cosmology in the Lagrangian
framework by performing spatial averages of Eulerian kinematical 
fields such as the rate of expansion $\theta:=\nabla_x\cdot\bm{v}$ 
of the fluid flow\cite{buchertehlers,ehlersbuchert}. 
They have found that the background introduced by 
the spatial averaging obeys the FRW cosmology under the appropriate 
assumption that the peculiar velocity field, which is defined as a 
deviation from a Hubble flow in the Eulerian picture, obeys  the
periodic boundary condition on a sufficiently large scale (see below). 

In this paper we modify  the approach by  Buchert \& Ehlers; we will  
work entirely within the Lagrangian framework.  
Namely, we divide the trajectory field into mean flow and deviation
field, and then take the spatial average of the Lagrange-Newton system 
in order to introduce the horizon scale background as well as the 
large-scale averaged trajectory field. 
We arrive at the same conclusion  as  that of Buchert \& Ehlers 
when averaged over the horizon scale.
Then, in order to separate the non-linear dynamics 
from the large-scale dynamics, we further separate the deviation field
into two parts, the averaged large-scale field and the rest. 
The evolution equation for the large-scale field is then obtained 
by averaging the local dynamical equation  over a large domain much 
 smaller
 than the  present horizon scale  in which the periodic boundary condition 
is applied  for the small-scale perturbations.
In this way we will obtain the evolution equations for the averaged large-scale
field and the local-scale field.  
The evolution equation for the local-scale field  naturally comes out
by subtracting the averaged evolution equation for the large-scale
field from the non-averaged equation. 
These equations couple to  each other in general situations. 
Therefore, we are able to  study  how  the smoothed large-scale structure 
is formed when the universe has non-linear structures on small scales 
as well as how the small-scale fluctuations grow 
in the surrounding environment.

This paper is organized as follows. In Section 2, we shall write down 
the basic equations needed in our considerations in the Lagrangian 
formalism developed by Buchert. In Section 3 we investigate the average 
properties of the equations derived in Section 2. We will have the
averaged FRW background under the periodic boundary condition over the
horizon scale. In Section 4, we develop the formalism to have
evolution equations for the large-scale and small-scale fluctuations 
where the results in Section 3 are used frequently. In general the 
evolution of the large-scale fluctuations is influenced by the
existence of the small-scale nonlinearity. We clarify the situations 
where the large-scale fluctuation 
behaves independently of  small-scale structures. 
In those cases the
evolution of the large-scale fluctuations 
 can be  described by the so-called ``truncated'' or ``optimized''
 Lagrangian perturbation theory which has  been originally 
 developed by many
 authors\cite{Coles,hamana,Melott94,Melott95,weiss} in order to 
avoid the shell-crossing problem of nonlinearity on small scales.  
Then it is also shown that the small-scale  dynamics 
is governed by the modified scale factor. 
The final section contains discussions and summary.
Throughout this paper, Latin indices take $1,2,3,$ respectively.

\section{Basic equations in the Lagrangian picture}\label{basicequations}
Let us start with the basic system of equations in Newtonian cosmology 
 describing  the motion of a self-gravitating pressureless fluid, 
so-called ``dust''. 
The dynamics of the fluid obeys the following familiar 
{\it Euler-Newton} system of equations in Newtonian hydrodynamics: 
{
\setcounter{enumi}{\value{equation}}
\addtocounter{enumi}{1}
\setcounter{equation}{0}
\renewcommand{\theequation}{\thesection.\theenumi\alph{equation}}
\begin{eqnarray}
&& \frac{\partial \rho}{\partial t}+\nabla_x\cdot \left(\rho
  \bm{v}\right)=0,  \label{eqn:continuity}\\
&& \frac{\partial \bm{v}}{\partial t} 
+(\bm{v}\cdot \nabla_x )\bm{v} = \bm{g},\label{eqn:euaccel} \\
&&\nabla_x \times \bm{g} =\mbox{{\bf 0}}, \label{eqn:irrotgrav}\\
&&  \nabla_x \cdot \bm{g} = -4\pi G \rho +\Lambda
 c^2,\label{eqn:poisson}
\end{eqnarray}\setcounter{equation}{\value{enumi}}}\noindent
where $\rho(\bm{x},t)$, $\bm{v}(\bm{x},t)$, and $\bm{g}(\bm{x},t)$
denote the fields of mass density, velocity, and gravitational
acceleration, respectively. The Poisson equation (\ref{eqn:poisson}) is 
extended including the cosmological constant for the sake of
generality. 
 
Following  the Lagrangian formulation  developed by
Buchert\cite{buchert89,buchertpre}, 
we concentrate on the integral curves $\bm{x}=\bm{f}(\bm{X},t)$ of the
velocity field 
$\bm{v}(\bm{x},t)$:
\begin{equation}
\frac{d \bm{f}}{dt}\left(=\dot{\bm{f}}\right) := \left.\frac{\partial \bm{f}}
{\partial t}\right|_X
=\bm{v}(\bm{f},t) ,\hspace{1.5cm} \bm{f}(\bm{X},t_I)\equiv \bm{X},
\label{eqn:map}
\end{equation}
where $\bm{X}$ denote the Lagrangian coordinates 
which label fluid elements, $\bm{x}$ are 
the positions of these elements in Eulerian space at time $t$, 
and $t_I$ is the initial time when  Lagrangian coordinates are defined. 

Then we can express 
the fields $\rho$, $\bm{v}$ and $\bm{g}$ 
in the Eulerian picture in terms of 
the Lagrangian coordinates $(\bm{X},t)$\cite{buchertpre,buchert93} from 
Eqs.(\ref{eqn:continuity}), (\ref{eqn:map}), and (\ref{eqn:euaccel}),
respectively:
{
\setcounter{enumi}{\value{equation}}
\addtocounter{enumi}{1}
\setcounter{equation}{0}
\renewcommand{\theequation}{\thesection.\theenumi\alph{equation}}
\begin{eqnarray}\label{eqn:lagfield}
&&\rho(\bm{X},t)=\frac{\mal{\rho}(\bm{X})}{J(\bm{X},t)},\label{eqn:lagmass}\\
&&\bm{v}(\bm{X},t)=\dot{f}(\bm{X},t),\label{eqn:lagvel}\\
&&\bm{g}(\bm{X},t)=\ddot{f}(\bm{X},t),\label{eqn:lagaccel}
\end{eqnarray}\setcounter{equation}{\value{enumi}}}\noindent
where $J$ is the determinant of the deformation field  $f_{i|j}$
(the vertical slash in the subscript denotes partial derivative with
respect to the Lagrangian coordinate $\bm{X}$) and the quantities with
$\mal{}$ 
such as $\mal{\rho}$ denote 
the quantities at the initial time $t_I$ henceforth, and 
we have used the fact $\mal{J}=1$. 
Thus, the continuity equation (\ref{eqn:continuity}) can 
be exactly integrated along the flow lines of the fluid elements in the 
Lagrangian picture\cite{buchert92,buchert93,buchertpre}. 
As a result, the dynamical variable in the Lagrangian picture 
 is only the  trajectory field $\bm{f}$. 
The equations (\ref{eqn:lagvel}) and
(\ref{eqn:lagaccel}) are similar to  point mechanics. The constraint
equations (\ref{eqn:irrotgrav}) and (\ref{eqn:poisson}) of the 
acceleration field $\bm{g}$ give us
the four evolution equations of the single dynamical field  $\bm{f}$ after 
the usual procedure in the Lagrangian formalism:
\begin{eqnarray}
 &&\epsilon_{abc}f_{i|a}f_{j|b}\ddot{f}_{j|c}=0,
  \label{eqn:rotationdiff} \\
 &&\frac{1}{2}
  \epsilon_{jab}\epsilon_{icd}f_{c|a}f_{d|b}\ddot{f}_{i|j}-\Lambda
  c^2 J=-4\pi G\mal{\rho}(\bm{X}), \
  \label{eqn:irrotationaldiff}
\end{eqnarray}
where we have used the mass conservation (\ref{eqn:lagmass}).  The set of
equations can be solved in principle for some initial conditions 
if we give the initial density field $\mal{\rho}\!\!(\bm{X})$ as a
source function.

The system of equations (\ref{eqn:lagfield}-c), (\ref{eqn:rotationdiff}),
and (\ref{eqn:irrotationaldiff}) is the so-called ``{\it Lagrange-Newton}''
system and equivalent to the Euler-Newton system 
as long as the mapping $\bm{f}_t:\bm{X}\mapsto \bm{x}$ 
is invertible. 
Buchert has solved the above set of equations perturbatively 
taking the solutions of FRW models as  
the zero-th order background solution 
(as discussed below)\cite{buchert92,buchert93,buchert94}.

For the purpose of  the later discussion and as an 
illustration of the Lagrangian formalism, we 
consider  the Eulerian vorticity field
$\omega_i=(1/2)\epsilon_{ijk}v_{k,j}$ and derive 
the {\it Kelvin's circulation  theorem} 
in the Lagrangian representation\cite{batcherlor,buchert92,buchertpre}. 
First, from (\ref{eqn:lagvel})  
we can rewrite the vorticity field $\bm{\omega}$ in terms of the
trajectory field $\bm{f}$ as 
\begin{equation}
\omega_i=\frac{1}{2J}\epsilon_{abc}f_{j|a}f_{i|b}\dot{f}_{j|c}.\label{eqn:euvor}
\end{equation}
To derive the theorem, we need Eq.(\ref{eqn:rotationdiff}). Multiplying 
the equation by the $h_{l,i}$, the inverse matrix of $f_{i|a}$, we can obtain 
\begin{equation}
\epsilon_{lbc}f_{j|b}\ddot{f}_{j|c}=0.
\end{equation} 
This equation can be  rewritten as 
\begin{equation}
\frac{d}{dt}\left(\epsilon_{lbc}f_{j|b}\dot{f}_{j|c}\right)=0, 
\end{equation}
so it can be integrated exactly along the trajectory field:
\begin{equation}
\frac{1}{2}\epsilon_{blc}f_{j|b}\dot{f}_{j|c}=\mal{\omega}_l,
\label{eqn:intomega}
\end{equation}
where we have used the initial condition
$\mal{\omega}_i=(1/2)\epsilon_{ijk}\dot{f}_{k|j}(\bm{X},t_I)$ 
from Eq.(\ref{eqn:euvor}). 
Finally, multiplying the above equation (\ref{eqn:intomega}) 
by the deformation field  $f_{i|l}/J$ and 
using Eq.(\ref{eqn:euvor}), we can obtain the following
Kelvin's circulation theorem along flow lines  
we are looking for:
\begin{equation}
\omega_{i}\!(\bm{X},t)=\frac{1}{J}f_{i|j}\mal{\omega}_{j}\!(\bm{X}).  
\label{eqn:vortheorem}
\end{equation}
The vorticity field in the Eulerian picture evolves according to the
above equation along the flow lines and is coupled to the density
enhancement, because the field is proportional to the inverse of 
the determinant 
of the deformation field  as the mass conservation equation 
(\ref{eqn:lagmass})\cite{buchert92}. This equation also means that 
if the initial vorticity field is zero at some  point, the vorticity 
field remains zero at any later time along the flow lines.
  Conversely, as the density field develops singularities
($J\rightarrow 0$), the vorticity field will blow up simultaneously 
even if the initial vorticity field is much smaller than the irrotational
part and is not zero\cite{buchert92}. Therefore, one should bear in mind 
that the vorticity field might play an important role 
in structure formation in the non-linear regime.

\section{Averaging Newtonian cosmologies}
\subsection{A Hubble  flow for a trajectory field}
\label{hubbleflow}
Before discussing the averaging problem, we first consider the 
properties  
of the trajectory field $\bm{f}$ in the 
Lagrange-Newton system derived 
in the previous section. 
We have reduced the description of the dynamics of any Eulerian 
field to the problem of finding the field of trajectories $\bm{f}$ as a
solution of the Lagrange-Newton system (\ref{eqn:rotationdiff}) and
(\ref{eqn:irrotationaldiff}). 
As in the Eulerian case, we are not able to write down any exact solution for
generic initial data $\mal{\rho}\!(\bm{X})$ 
without assuming a symmetry like plane or spherical symmetry. 
We may start with the simplest class of solutions,
the homogeneous-isotropic ones, and then move on to the 
treatment of inhomogeneities.

Those fluid motions which are locally
isotropic in the sense that, at any time and for each particle ${\cal
P}$, there exists a neighborhood on which the field of velocities
relative to ${\cal P}$ is invariant under all rotations about ${\cal
P}$, are given by the following form with our  choice
(\ref{eqn:map}) of Lagrangian  coordinates: 
\begin{equation}
 \bm{x}=\bm{f}_H(\bm{X},t)=\frac{a(t)}{\mal{a}}\bm{X},
  \hspace{3em\mal{a}}:=a(t_I),
  \label{eqn:homotraj}
\end{equation}
if we conventionally put $\bm{f}_H(\mbox{{\bf 0}},t)=\mbox{{\bf
0}}$. 
Such a flow is a well-known Hubble flow.  Inserting this
ansatz into the Lagrange-Newton system (\ref{eqn:rotationdiff}),
(\ref{eqn:irrotationaldiff}) and the mass conservation (\ref{eqn:lagmass})
yield the usual Friedmann  equations\cite{buchert89,buchertpre}
\begin{eqnarray}
 &&\frac{\ddot{a}(t)}{a(t)}=-\frac{4\pi
  G}{3}\frac{\mal{\rho}_b\mal{a}^3}{a^3(t)}+\frac{\Lambda c^2}{3},
  \label{eqn:lagfriedac}\\
 && \rho_b(t)=\frac{\mal{a}^3}{a^3(t)}\mal{\rho}_b. \label{eqn:laghomodens}
\end{eqnarray}
Thus, the quantity $a(t)$ agrees with  the scale factor in FRW
cosmology. 
It should be remarked that the assumption of homogeneous 
and isotropic matter flow (\ref{eqn:homotraj})  makes the 
initial density independent of $\bm{X}$
via the equation (\ref{eqn:irrotationaldiff}):
$\mal{\rho}\!(\bm{X})\equiv\mal{\rho}_b=\mbox{constant}$. In other words, the
existence of the fluctuation for the initial density field 
produces no longer a Hubble flow such as (\ref{eqn:homotraj}).  
We may use it to integrate Eq.(\ref{eqn:lagfriedac})
yielding Friedmann's differential equation:
\begin{equation}
 \frac{\dot{a}^2}{a^2}+\frac{{\cal K}c^2}{a^2}=\frac{8\pi G}{3}
  \frac{\mal{\rho}_b\mal{a}^3}{a^3}+\frac{\Lambda c^2}{3};\hspace{2em}{\cal K}=
  \mbox{const.}\label{eqn:lagfried}
\end{equation}
where ${\cal K}$ is the constant of integration mathematically and can
be regarded as the curvature parameter of the FRW model. 
Naturally,  
Eq.(\ref{eqn:euvor}) gives 
\begin{equation}
\bm{\omega}=\mbox{{\bf 0}}.
\end{equation}
Thus the assumption (\ref{eqn:homotraj}) for the trajectory field 
produces the standard Friedmann 
cosmologies. 

\subsection{Averaged properties of the Lagrange-Newton system in 
an inhomogeneous universe}
We now consider the trajectory field when there exist inhomogeneities in 
the universe. In the application of Lagrangian theory to 
the averaging problem in cosmology, we examine the behavior of 
 some spatially compact domain ${\cal D}(t)$ on the Eulerian  space   
occupied by the fluid elements, which corresponds to
the initial domain $\mal{{\cal D}}$ of the Lagrangian coordinates
via the mapping $\bm{f}_t:\bm{X}\mapsto \bm{x}$.  
 
For our purpose, 
we set the average flow in the form of a Hubble flow with scale 
factor $a_{\cal D}$ not necessarily equal to  
$a(t)$ and 
define the  (not necessarily small) deviation
field $\bm{P}$ from the average flow without loss of generality,
so that the full trajectory field $\bm{f}$ of an inhomogeneous model reads:
\begin{eqnarray}
&& \bm{f}(\bm{X},t):=\bm{f}^{{\cal D}}_{H}\!(\bm{X},t)+\bm{P}(\bm{X},t)
=\frac{a_{\cal D}(t)}{\mal{a}_{\cal D}}\bm{X}+\bm{P}(\bm{X},t);\hspace{2em} 
\bm{P}(\bm{X},t_I):=\mbox{{\bf 0}},
  \label{eqn:inhomotraj}
\end{eqnarray}
Thus, for the sake of convenience, we here 
start with  the trajectory field 
in the form of Zel'dovich type solution\cite{Zeldovich}.
However, it should be noted 
 that we can define the full trajectory field  by
Eq.(\ref{eqn:inhomotraj}) for any fluid elements 
in an arbitrary initial domain $\mal{{\cal D}}$, and 
the scale factor depends naturally on the chosen domain. 
We also remark that since we have imposed no condition on
the deviation field $\bm{P}$, 
this consideration is applied to  the non-linear situations when the
density 
contrast field may be larger than unity.  

Using  the full trajectory field (\ref{eqn:inhomotraj}),
we obtain by a little computation the
following 
expression for the determinant $J$ of deformation field $f_{i|j}$:
\begin{eqnarray}
 &&J=\frac{1}{3!}\epsilon_{abc}\epsilon_{ijk}f_{a|i}f_{b|j}f_{c|k}\nonumber \\
 &&\hspace{1em}=\frac{a_{{\cal D}}^3}{\mal{a}_{{\cal D}}^3}+\frac{a_{{\cal
  D}}^2}{\mal{a}_{{\cal D}}^2}P_{i|i}+\frac{1}{2}\frac{a_{{\cal D}}}{\mal{a}_{{\cal 
  D}}}\left(P_iP_{j|j}-P_{j}P_{i|j}\right)_{|i}
  +\frac{1}{6}\left[P_i\left(P_{j|j}P_{k|k}-P_{j|k}P_{k|j}
   \right)+2P_j\left(P_{i|k}P_{k|j}-P_{k|k}P_{i|j}\right)\right]_{|i}
  \nonumber \\
 &&\hspace{1em}:=\frac{1}{\mal{a}_{{\cal D}}^3}\left(a_{{\cal D}}^3(t)+\tilde{{\cal 
  J}}_{i|i}(\bm{X},t)\right), \label{eqn:averdeterm}
\end{eqnarray}
where   
\begin{equation}
 \tilde{{\cal J}}_i:=\mal{a}_{{\cal D}}a_{{\cal
  D}}^2P_i+\frac{\mal{a}_{{\cal D}}^2}{2}a_{{\cal D}}
  \left(P_iP_{j|j}-P_{j}P_{i|j}\right)+
  \frac{\mal{a}_{{\cal D}}^3}{6}\left[P_i\left(P_{j|j}P_{k|k}-P_{j|k}P_{k|j}
   \right)+2P_j\left(P_{i|k}P_{k|j}-P_{k|k}P_{i|j}\right)\right].
\end{equation}
Note that  the second term on the right-hand-side of
Eq.(\ref{eqn:averdeterm}) is expressed by the divergence of 
the vector $\tilde{\bm{{\cal J}}}(\bm{X},t)$ with
respect to the Lagrangian coordinates and   
it  is defined in terms of the deviation displacement 
vector $\bm{P}$. 
Since the volume elements at $t$ and $t_I$ are related by
$d^3x={\cal J}d^3X$, 
we use the above equation to rewrite the volume $V_{{\cal D}}(t)$ of
the domain
${{\cal D}}(t)$ of the fluid in the following form: 
\begin{eqnarray}
 V_{{\cal D}}(t):=\int_{{\cal D}(t)}\!d^3x
  &=&\int_{\mal{{\cal D}}}\!d^3X {\cal J}(\bm{X},t) \nonumber \\
 &=&\frac{1}{\mal{a}_{{\cal D}}^3}\left(a_{{\cal D}}^3
  +\kaco{\tilde{{\cal J}}_{i|i}}_{\mal{{\cal 
  D}}}\right) \mal{V}_{{\cal D}}\nonumber \\
 &=& \frac{a_{{\cal D}}^3}
{\mal{a}_{{\cal D}}^3}\mal{V}_{{\cal D}}+\frac{1}{\mal{a}_{{\cal 
  D}}^3}
  {\displaystyle \int_{\partial \mal{{\cal D}}}\!d\bm{S}_X\cdot
\tilde{\bm{{\cal J}}}(\bm{X},t)},  \label{eqn:volrel}
\end{eqnarray}
where 
we have applied Gau\ss's theorem to transform the volume integral 
to a surface integral over the boundary $\partial
\!\mal{{\cal D}}$ of the initial domain:
\[
 \int_{\mal{{\cal D}}}\!d^3X \nabla_X\cdot\tilde{\bm{{\cal
J}}}=\int_{\partial \mal{{\cal D}}}\!d\bm{S}_X\cdot\tilde{\bm{{\cal J}}}.
\]
In the derivation of Eq.(\ref{eqn:volrel}), 
$\kaco{\ldots}_{\mal{{\cal D}}}$ denotes
the spatial average of a tensor field over
the initial domain $\mal{{\cal D}}$, and we regard
$a_{{\cal D}}(t)$ as the scale factor of that  domain ${\cal D}(t)$ in
the Eulerian space. 
The quantity  $\mal{V}_{{\cal D}}$ denotes the volume of initial domain
$\mal{{\cal D}}$ considered,
\begin{equation}
\mal{V}_{\cal D}:=\int_{\mal{\cal D}}d^3X. 
\end{equation}
A note of caution is in order: 
Buchert \& Ehlers\cite{buchertehlers} have used  a  
domain dependent 
scale factor $a_{\cal D}$ defined by 
 $V_{\cal D}\equiv 
a^3_{\cal D}(t)$, but Eq.(\ref{eqn:volrel}) means that  
such a scale factor does not agree with our scale factor 
defined by Eq.(\ref{eqn:inhomotraj}).  Thus, these two concepts 
are different in the following respect: if we would impose 
{\it periodic boundary conditions} for $\bm{P}$ on the domain, then our 
scale factor reduces to the standard FRW scale factor on that domain 
as discussed below. 

Likewise, the spatial average of the density field
(\ref{eqn:lagmass}) over the domain ${\cal D}$ in Eulerian space 
at any time $t$ may be calculated as follows.
\begin{eqnarray}
  \kaco{\rho}_{{\cal D}}\!(t)
&=&\frac{1}{V_{\cal
  D}}\int_{{\cal D}(t)}\!d^3x\frac{\mal{\rho}(\bm{X})}{{\cal J}(\bm{X},t)}
  \nonumber \\
 &=&\frac{1}{V_{{\cal D}}}\int_{\mal{{\cal D}}}\!d^3X\mal{\rho}\nonumber 
  \\
 &=&\frac{\mal{V}_{{\cal D}}}{V_{{\cal D}}(t)}\langle\mal{\rho}\rangle_{\mal{{\cal 
  D}}}\nonumber \\
 &=& \frac{\mal{a}_{{\cal D}}^3}{a_{{\cal D}}^3+\left(1/\mal{V}_{{\cal
  D}}\right){\displaystyle \int_{\partial \mal{{\cal D}}}\!d\bm{S}_X\cdot\tilde{\bm{{\cal
  J}}}}}\langle\mal{\rho}(\bm{X})\rangle_{\mal{{\cal D}}}.\label{eqn:aveeudens}
\end{eqnarray}

Next we consider the average of
the dynamical equation (\ref{eqn:irrotationaldiff})
of the trajectory field. Likewise, the first term on the left-hand-side of
Eq.(\ref{eqn:irrotationaldiff}) can
be rewritten by inserting Eq.(\ref{eqn:inhomotraj}) in the following
form:
\begin{eqnarray}
 \frac{1}{2}\epsilon_{abc}\epsilon_{ijk}f_{a|i}f_{j|b}\ddot{f}_{k|c}&=&
  \frac{1}{\mal{a}_{{\cal D}}^3}\left(3a_{{\cal D}}^2\ddot{a}_{{\cal D}} 
  +{\cal Q}_{i|i}(\bm{X},t)\right),\label{eqn:irrotaver}
\end{eqnarray}   
where 
\begin{eqnarray}
 && {\cal Q}_i(\bm{X},t):= \mal{a}_{{\cal D}}a_{{\cal
   D}}^2\left(\ddot{P}_i+2\frac{\ddot{a}_{{\cal D}}}{a_{{\cal D}}}P_i\right) 
\nonumber \\
&&\hspace{5em}+\mal{a}^2_{{\cal D}}a_{{\cal
   D}}\left(\ddot{P}_iP_{j|j}-\ddot{P}_jP_{i|j}\right)
   +\frac{\mal{a}^2_{{\cal D}}}{2}\ddot{a}_{{\cal
   D}}\left(P_iP_{j|j}-P_jP_{i|j}\right) \nonumber \\
 &&\hspace{5em}+\frac{\mal{a}_{{\cal D}}^3}{2}
   \left[\ddot{P}_i\left(P_{j|j}P_{k|k}-P_{j|k}P_{k|j}\right)
    + 2\ddot{P}_j\left(P_{i|k}P_{k|j}-P_{i|j}P_{k|k}\right)\right],
\end{eqnarray}
and we again remark that
the second term on the right-hand-side of
Eq.(\ref{eqn:irrotaver}) can also  be expressed
by the divergence of
the vector $\bm{{\cal Q}}(\bm{X},t)$.
Thus, we can get the local evolution equation for the domain dependent
scale factor $a_{\cal D}$ and the displacement vector $P_i$:
\begin{eqnarray}
&&3a_{\cal D}^2\ddot{a}_{\cal D}+4\pi G\mal{\rho}\!\!(\bm{X})
\mal{a}_{\cal D}^3-\Lambda c^2a_{\cal D}^3=-{\cal Q}_{i|i}+
\Lambda c^2{\cal J}_{i|i}.\label{eqn:originaleq}
\end{eqnarray}
Averaging over the initial domain $\mal{{\cal D}}$ of
the above equation (\ref{eqn:originaleq}) leads to the following equation:
\begin{equation}
 3\frac{\ddot{a}_{{\cal D}}}{a_{{\cal D}}}+
  4\pi G\frac{\langle\mal{\rho}(\bm{X})\rangle_{\mal{{\cal D}}}
  \mal{a}_{{\cal D}}^3}{a^3_{{\cal D}}}-\Lambda c^2=
  \frac{1}{a^3_{{\cal
  D}}\mal{V}_{{\cal D}}}\int_{\partial \mal{{\cal D}}}\!d\bm{S}_X\cdot
  \left(-\bm{{\cal Q}}+\Lambda c^2 \tilde{\bm{{\cal J}}}\right), 
\label{eqn:nbackreaction}
\end{equation}
where we again used Gau\ss's theorem.
If the local equation (\ref{eqn:originaleq})
 and the averaged equation (\ref{eqn:nbackreaction}) 
can be solved simultaneously, 
the domain dependent scale factor $a_{\cal D}(t)$  
and the local displacement vector $P_i$ are  obtained in principle, 
respectively.
This averaged equation can also  be interpreted  as 
a standard Friedmann equation for the ``effective mass density'' 
$\rho_{\mbox{{\scriptsize eff}}}$\cite{buchertehlers}, which is here 
 defined by
\begin{equation}
 4\pi G\rho_{\mbox{{\scriptsize eff}}}(t):=
  4\pi G \frac{\langle\mal{\rho}\rangle_{\mal{{\cal D}}}\mal{a}_{\cal D}^3}
{a^3_{{\cal D}}}+
  \frac{1}{a^3_{{\cal D}}\mal{V}_{{\cal D}}}\int_{\partial \mal{{\cal
  D}}}\!d\bm{S}_X\cdot\left(\bm{{\cal Q}}-\Lambda c^2\tilde{\bm{{\cal
		       J}}}\right),\label{eqn:effectmassfried}
\end{equation}
where the first term on the right-hand-side decreases clearly 
 in proportion to  $a_{\cal D}^{-3}$. 
The equation (\ref{eqn:nbackreaction}) shows that inhomogeneities
have an accelerating effect on the  expansion rate $\dot{a}_{{\cal
D}}/a_{{\cal D}}$ of  the average flow, if the term 
$\int_{\partial \mal{\cal
D}}\!d\bm{S}_X\cdot (-\bm{{\cal Q}})$ on the right-hand-side dominates
the other  terms and is positive.
Namely, this shows that the evolution of the domain dependent scale 
factor $a_{{\cal D}}$ does in fact depend on the chosen domain, 
that is, the averaged expansion will be different
from  the usual Friedmann laws (\ref{eqn:lagfriedac}) 
if the averages involving $\bm{{\cal Q}}$ 
and $\tilde{\bm{{\cal J}}}$ do not vanish.

Next, we consider the property of 
the average of the vorticity field $\bm{\omega}$. 
Similarly, performing 
the average of Eq.(\ref{eqn:vortheorem}) 
over the domain ${\cal D}$ on the Eulerian space (not the Lagrangian space), 
we can obtain 
\begin{eqnarray}
\kaco{\omega_i}_{{\cal D}}=\frac{\mal{a}_{{\cal
  D}}^3}{a_{{\cal D}}^3(t)+\left(1
  /\mal{V}_{{\cal D}}\right){\displaystyle \int_{\partial \mal{\cal D}}\!\!d\bm{S}_X
\cdot\tilde{\bm{{\cal
J}}}}\!(\bm{X},t)}\left[\frac{a_{{\cal D}}}{\mal{a}_{\mal{{\cal D}}}}
\langle\mal{\omega}_i\rangle_{\mal{{\cal
 D}}}+\frac{1}{\mal{V}_{{\cal D}}}\int_{\partial\mal{{\cal D}}}\!\!dS_{Xj}\left(
 P_{i}\mal{\omega}_j\right)\right].\label{eqn:avervor}
\end{eqnarray}
Noting that the quantity $\mal{\omega}_i$ is divergence-less by 
definition, namely $\mal{\omega}_{i|i}=(1/2)\epsilon_{ijk}\mal{v}_{k|ji}=0$, 
 we have  again used  Gau\ss's theorem in rewriting the second term 
on the right-hand-side. 
The equation (\ref{eqn:avervor}) means that if  
$\mal{\omega}_i$ vanishes at every point on  the initial 
hypersurface, it leads to $\kaco{\omega_i}_{\cal D}=0$ at any time.  

Thus, the average properties of the Lagrange-Newton system do not necessarily 
agree with the FRW cosmologies in a general inhomogeneous universe. 
However, based on the observation of extreme isotropy of the CMB, 
we expect that the universe is almost 
isotropic and homogeneous on a sufficiently large scale.
We consider how this fact is expressed mathematically in terms of
averaged variables. 

As discussed above, 
we could define  the displacement vector $\bm{P}$
as   representing  the deviation from
the mean flow generated by the inhomogeneities.
The resultant equation (\ref{eqn:nbackreaction}) then shows
us how this field $\bm{P}$ determines the backreaction 
on the scale $a_{{\cal D}}(t)$ from Friedmann's law. 
Note that the backreaction terms are expressed by the surface integrals
 over the boundary of the initial domain
$\partial\!\!\mal{{\cal D}}$. 
As already noted,  if we employ the periodic boundary condition for the
deviation vector field $\bm{P}$ on some sufficiently large scale
${\cal D}_p$, the backreaction terms in 
Eq.(\ref{eqn:nbackreaction}) are exactly zero:
\begin{equation}
 3\frac{\ddot{a}_{{\cal D}_p}}{a_{{\cal D}_p}}+4\pi G\frac{ \mal{\rho}_b
  \mal{a}_{{\cal D}_p}^3}{a_{{\cal D}_p}^3}-\Lambda c^2=0,
\label{eqn:averfriedmannacce}
\end{equation}
with
\begin{equation}
 \mal{\rho}_b\equiv\langle\mal{\rho}(\bm{X})\rangle_{\mal{{\cal D}}_p}=\mbox{constant}.
  \label{eqn:averinitdens}
\end{equation}
Similarly, Eq.(\ref{eqn:aveeudens}) gives the background density at an
arbitrary time: 
\begin{equation}
 \rho_b(t)\equiv \kaco{\rho(\bm{X},t)}_{{\cal
  D}_p(t)}=\frac{\mal{\rho}_b\mal{a}_{{\cal D}_p}^3}{a_{{\cal D}_p}^3(t)}.
\label{eqn:averdensity}
\end{equation}
We thus obtain the usual definitions of the homogeneous
background density field  (\ref{eqn:averinitdens}) and
(\ref{eqn:averdensity}) using the spatial averaging.
Since we have not restricted ourselves to the perturbative situation
where the deviation vector $\bm{P}$ is infinitesimally small,
this discussion is always valid for  non-linear situations 
under the periodic boundary condition. 

Under the same periodic assumption, Eq.(\ref{eqn:avervor}) 
 gives us 
the following form as the average of the Eulerian vorticity field: 
\begin{equation}
\kaco{\omega_i}_{{\cal D}_p}=\frac{\mal{a}_{{\cal D}_p}^2}{a_{{\cal
 D}_p}^2}\langle\mal{\omega_i}\rangle_{\mal{{\cal D}}_p}
\end{equation}
with the scale factor $a_{{\cal D}_p}$ defined 
by Eq.(\ref{eqn:averfriedmannacce}).  
Thus the averaged vorticity field  decays as $\propto
a_{{\cal D}_p}^{-2}$ in the expanding universe, which is analogous to  
the Newtonian linearized theory\cite{peebles}.  
In other words, even if the initial global averaged
 value $\langle\mal{\omega_i}\rangle_{\mal{{\cal D}}_p}$
is not zero, we can safely ignore  the global averaged 
vorticity field. 

The COBE microwave background measurement suggests that
the power spectrum of the density fluctuation field
has a positive slope on large scales, supporting the
assumption of large-scale homogeneity. 
This suggests that even if we do not employ the periodicity 
over the horizon scale, the flux of $\bm{{\cal Q}}$ and 
$\tilde{\bm{{\cal J}}}$ in Eq.(\ref{eqn:nbackreaction}) 
through the boundary of the averaging domain with a sufficiently 
large volume may be negligible.  
Therefore,  we may conclude that the backreaction on the global expansion rate
becomes  zero and the equations of the background model
introduced by the spatial average of an inhomogeneous universe over
the horizon scale obey Friedmann's laws
(\ref{eqn:lagfriedac}) and (\ref{eqn:laghomodens})
in Newtonian cosmology, even when the universe has locally nonlinear
structures ($\delta\gg 1$) on some small scales. 

We may proceed to solve the set of equations (\ref{eqn:rotationdiff})
and (\ref{eqn:originaleq}) perturbatively by taking the solutions of
the Friedmann's laws as the zeroth order approximation for the scale
factor to construct a locally  inhomogeneous universe\cite{buchert92,buchert94}. 
This approach is certainly useful, but it is not easy to see 
the effect of large-scale structure on the small-scale nonlinear dynamics. 
To see this explicitly we will take a new approach in the next section.

\section{Hybrid Lagrangian theory for Newtonian gravitational 
instability}
   
In 1970 Zel'dovich\cite{Zeldovich}  
found the so-called Zel'dovich approximations to 
describe the large-scale structure
formation; it has been derived in the gravitational context 
by Buchert\cite{buchert89} and it can also be obtained as a 
particular first-order solution in the Lagrangian perturbation theory 
 developed by
Buchert\cite{buchert92,buchert93,buchert94,buchertpre} and many
other authors\cite{bouchet,Coles,sahni}.  
It has been shown that Zel'dovich  approximation gives indeed a very good
approximation and has been used up to the quasi non-linear regime 
to reproduce the observed filament-like and pancake-like  
 pattern of the large-scale structure beyond several megaparsecs.

The numerical simulation based on the approximation has an advantage 
over the N-body simulation such that it is able to simulate 
relatively large domains with relatively small memories.  
However, the Zel'dovich approximation also  has a disadvantage.
Namely,  it cannot reproduce the non-linear structure formation 
on small scales once the shell crossing occurs. This is called the 
 {\it  shell crossing problem}.  
To avoid this, the {\it truncated} or {\it optimized} 
Lagrangian perturbation approaches have been
developed by many authors\cite{Coles,hamana,Melott94,Melott95,weiss} where 
one smoothes out the initial small-scale 
fluctuations such that the shell-crossings occur at about the present time. 
This will be a good approximation for the evolution of the 
large-scale structure, if one can show that the evolution does not very
much depend upon the behavior of small-scale  nonlinear dynamics. 
However, this expectation has not yet been explicitly proven or, put in 
another way, it has not yet been clarified under what sort of
situations this expectation is valid. 
If this is proven, there may be some hope to have a hybrid way to
describe the  inhomogeneous universe such that the large-scale
structure is described by Zel'dovich's approximation or some improved
version of it and the local small-scale dynamics is described by other
method such as N-body simulation or some effective theory of nonlinear
structure formation.

This is what we wish to develop in this section. We do this by
dividing the deviation field into two parts, the
averaged large-scale part and the small-scale part. 
In the following, we consider the Einstein-de Sitter
 universe ($\Lambda={\cal K}=0$) for the sake of simplicity, 
and the scale factor is normalized as 
\begin{equation}
\mal{a}_{\cal D}=1.
\end{equation}
Namely, we set the scale factor so that 
 the comoving coordinates agree with the Lagrangian
coordinates.

\subsection{The division of the deviation field into the large-scale
 part and the small-scale part}

Let us start writing the trajectory field as 
\begin{equation}
f_i(\bm{X},t)\equiv f^H_i(\bm{X},t)+P_i(\bm{X},t;\lambda)
= a(t)X_i+P_i(\bm{X},t;\lambda),
\label{eqn:inhomo}
\end{equation}
where we have explicitly included the wavelength ($\lambda$)
dependence  in the deviation field  
$P_i$. Note that $\lambda$ here denotes the wavelength of 
the initial density fluctuation field in the comoving coordinates 
as explained later. 
Here the quantity $a(t)$  is the scale factor defined in the previous
section: $a=a_{{\cal D}_p}$, (where we have omitted the subscript of
the quantity $a_{{\cal D}_p}$  for simplicity).   
Namely, the scale factor  is obtained by averaging of the Lagrange-Newton
system over the horizon scale.  
Following the results in the previous section, 
we assume henceforth that it obeys Friedmann's law, so  
the deviation field $P_i$ obeys 
the periodic boundary condition 
on the present horizon scale:
\begin{equation}
\kaco{P_{i|j}(\bm{X},t;l<\lambda<L_H)}_{\mal{V}_H}=0.
\label{eqn:horizonperiod}
\end{equation}
The wavelength of the initial density fluctuations has the lower
cutoff, because we deal with a collisionless gravitational system 
like dark matter. Below the lower cutoff this description is not valid
since the baryonic gaseous pressure and the effective 
pressure due to the velocity dispersion of the collisionless system 
may become important. We will not consider such small scales. 
We have also the upper cutoff which will be the horizon scale 
$L_H$ because we assumed that the deviation field $P_i$ obeys 
the periodic boundary condition 
(\ref{eqn:horizonperiod}) on the horizon scale $L_H$. 
We will not explicitly write down these cutoff lengths hereafter.

By substituting the ansatz (\ref{eqn:inhomo}) into 
 Eq.(\ref{eqn:irrotationaldiff}),  we can obtain
 Eq.(\ref{eqn:originaleq}), or more explicitly as
\begin{eqnarray}
&&3a^2\ddot{a}+a^2\ddot{P}_{i|i}+2a\ddot{a}P_{i|i}
+a\left(P_{i|i}\ddot{P}_{j|j}-P_{i|j}\ddot{P}_{j|i}\right)
\nonumber \\
&&\hspace{8em}+\frac{\ddot{a}}{2}\left(P_{i|i}P_{j|j}-P_{i|j}
P_{j|i}\right)+\frac{1}{2}\epsilon_{ijk}\epsilon_{abc}
P_{i|a}P_{j|b}\ddot{P}_{k|c}
=-4\pi G\mal{\rho}(\bm{X};l<\lambda<L_H)\label{eqn:originaleq2}.
\end{eqnarray}

Now we introduce  the averaged vector field  $\pla_i$ 
by using the spatial average of the full deviation field $P_i$ 
over the large-scale domain  $\mal{{\cal D}}_L\!\!(\bm{X})$ 
(which is still much smaller than the horizon scale)
at some point $\bm{X}$ in the Lagrangian coordinates:
\begin{eqnarray}
p^>_{i}(\bm{X},t;\lambda\gtrsim L)
&\equiv&\frac{1}{\mal{V}_L}\int_{\mal{{\cal
D}_L}(X)}d^3X^\prime 
P_{i}(\bm{X}^\prime,t;\lambda)\nonumber \\
&:=&\frac{1}{\mal{V}_L}\int
d^3X^\prime P_{i}(\bm{X}^\prime,t;\lambda)
W(\bm{X}-\bm{X}^\prime;L), 
\label{eqn:mediumdis}
\end{eqnarray}
where  $L$ is an artificial cutoff length and
\begin{equation}
\mal{V}_L:=
\int_{\mal{{\cal D}_L}(X)}\!\!\!d^3X=\int d^3X^\prime 
W(\bm{X}-\bm{X}^\prime;L), 
\end{equation}
and $W(\bm{X}-\bm{X}^\prime;L)$ is a filter 
function characterized by the smoothing length $L$.
For example, for Gaussian filtering on scale $L$ the filter function is 
\begin{equation}
W_G(r;L)=\frac{1}{(2\pi)^{3/2}L^3}\mbox{exp}[-r^2/(2L^2)].
\end{equation}
From Eq.(\ref{eqn:inhomotraj}) the initial condition is imposed on the $\pla_i$:
\begin{equation}
\pla_{i}(\bm{X},t_I)=0.\label{eqn:larginit}
\end{equation}
We have assumed  that, since the smoothing method with 
the scale $L$
is used for the definition of $\pla_i$, its valid 
wavelength range  has about the length $L$ as 
a lower cutoff length.  Thus, we expect  that 
the cutoff scale length $L$ is large enough so that 
the $p^>_i$ describes the evolution of  fluctuations with 
the characteristic length larger than $L$ in the linear 
regime still at the present time $t_0$, that is,
$|p^>_{i|j}(\bm{X},t_0)|\ll  
a(t_0)$ is satisfied according to the Lagrangian perturbation theory. 
It should be noted that 
the linear regime in the Lagrangian picture does not mean 
$\delta\ll 1$ because the density field can be exactly solved.
In this situation, 
we can safely say that the cutoff length $L$ 
is supposed to be in the range $l\ll L \ll L_H$ in the realistic universe.
In the following, we consider  only the lowest order in $\pla_i$ by
means of this assumption.

Simultaneously, since the original displacement vector $P_i$ 
describes the gravitational instability of fluctuations 
with scale larger than $l$ and smaller than $L_H$, 
the definition (\ref{eqn:mediumdis}) of $\pla_i$ allows us to 
define the small-scale displacement vector $\psm_i$
 which is supposed to describe 
non-linear structure formation on small scales.
Namely, the full displacement vector $P_i$ can be written as
\begin{equation}
P_i(\bm{X},t;\lambda)\equiv p^>_i(\bm{X},t;L\lesssim\lambda)
+p^<_i(\bm{X},t;\lambda\lesssim L). \label{eqn:fulltraject}
\end{equation}
The initial condition for $\psm_i$ is
\begin{equation}
\psm_i(\bm{X},t_I)=0.
\end{equation}
We expect that $p^<_i$ describes the evolution of 
the initial density fluctuation field  with  much smaller 
length  than our cutoff length $L$.

Next, we shall derive the evolution equations for the large-scale 
displacement vector $p^>_i$ and the small-scale 
displacement vector $p^<_i$, respectively. 
Before doing this we present a basic equation which is obtained by 
inserting the ansatz (\ref{eqn:fulltraject}) into 
Eq.(\ref{eqn:originaleq2}).
\begin{eqnarray}
&&3a^2\ddot{a}+a^2\ddot{p}^>_{i|i}+a^2\ddot{p}^<_{i|i}
+2a\ddot{a}p^>_{i|i}+2a\ddot{p^<}_{i|i}+
a\left(p^>_{i|i}\ddot{p}^<_{j|j}-p^>_{i|j}\ddot{p}^<_{j|i}
\right)+a\left(\ddot{p}^>_{i|i}p^<_{j|j}-\ddot{p}^>_{i|j}
p^<_{j|i}\right)\nonumber \\
&&+\ddot{a}\left(p^>_{i|i}p^<_{j|j}-p^>_{i|j}p^<_{j|i}\right)
+p^>_{i|i}\left[p^<_j\ddot{p}^<_{k|k}
-p^<_k\ddot{p}^<_{j|k}\right]_{|j}
+p^>_{i|j}\left[2p^<_{j|k}\ddot{p}^<_{k|i}-p^<_{k|k}\ddot{p}^<_{j|i}
-\ddot{p}^<_{k|k}p^<_{j|i}\right]\nonumber \\
&&+\frac{1}{2}\ddot{p}^>_{i|i}\left[p^<_{j}p^<_{k|k}
-p^<_kp^<_{j|k}\right]_{|j}
+\ddot{p}^>_{i|j}\left[\left(\psm_{j}\psm_{k|i}\right)_{|k}
-\left(\psm_{k|k}\psm_{j}\right)_{|i}\right]\nonumber \\
&&+a\left(\psm_{i}\ddot{p}^<_{j|j}-\psm_j
\ddot{p}^<_{i|j}\right)_{|i}
+\frac{\ddot{a}}{2}\left(\psm_{i}\psm_{j|j}
-\psm_j\psm_{i|j}\right)_{|i}
+\frac{1}{2}\left[\ddot{p}^<_i\left(\psm_{j|j}\psm_{k|k}
-\psm_{j|k}\psm_{k|j}\right)+2\ddot{p}_j\left(\psm_{i|k}\psm_{k|j}
-\psm_{i|j}\psm_{k|k}\right)\right]_{|i}\nonumber \\
&&\hspace{14em}
+O\left((\pla)^2\right)=-4\pi G\mal{\rho}(\bm{X};l<\lambda<L_H).
\label{eqn:fullevoleq}
\end{eqnarray}
In the above derivation, for example, we have  used  the 
results such as 
\[
\psm_{i|i}\psm_{j|j}-\psm_{i|j}\psm_{j|i}=\left[\psm_i\psm_{j|j}
-\psm_{j}\psm_{i|j}\right]_{|i}.
\]
Note that we have kept only linear order in $p^>_i$ and 
full order in $p^<_i$. 
It should be also noted that only the third term in the second line 
on the left-hand-side of Eq.(\ref{eqn:fullevoleq}) cannot be  
expressed in the form of a divergence of the vector which consists of 
$\psm_i$.  
In the discussion below, the large-scale 
transverse mode is  omitted  in the sense that  
 we use only one of the equations 
(\ref{eqn:rotationdiff}) and (\ref{eqn:irrotationaldiff}), namely
(\ref{eqn:fullevoleq}). For the smoothed large-scale field $\pla_i$, 
this may be a good approximation, because assuming that 
the initial vorticity field is negligible compared with the initial 
irrotational flow
results in vanishing of  the transverse part at a later time as explained 
in Appendix A (this may be a reasonable 
assumption based on the linearized theory\cite{peebles}), 
but there are transverse parts in the Lagrangian space in the {\it nonlinear} 
situation even for $\omega^i=0$. Buchert \& Ehlers\cite{buchert93} derived 
 the transverse solutions of 
Eqs.(\ref{eqn:continuity}-d) 
for the second-order transverse and irrotational solutions in a general case.

\subsection{The evolution equation for the large-scale structure 
formation}

First, we consider the evolution equation for the averaged 
large-scale field $\pla_i$.  
Since we have assumed that the $\psm_i$ describe the behavior 
of fluctuations with the characteristic scale much smaller than
$L$, we introduce the following spatial averaging method  
on the scale $L^\prime$ much smaller than $L$ and much larger than $l$:
$l\ll L^\prime\ll L$. Namely, the averaging is defined by 
\begin{equation}
\mbox{(1)}\hspace{3em} \kaco{g}_{L^\prime}(\bm{X},t)\equiv 
\frac{1}{\mal{V}_{L^\prime}}\int_{\mal{\cal D}_{L^\prime}(X)}
\hspace{-1.5em}d^3X^\prime
g(\bm{X}^\prime,t):=\frac{1}{\mal{V}_{L^\prime}}\int
d^3X^\prime g(\bm{X}^\prime,t) W(\bm{X}-\bm{X}^\prime;L^\prime).
\label{eqn:rule1}
\end{equation}
Furthermore, within the averaging volume $\mal{V}_{L^\prime}$, 
we can safely neglect the spatial gradient of $\pla_i$ 
because from Eq.(\ref{eqn:mediumdis}) 
the $\pla_i$ are defined by the averaging on 
the volume $\mal{V}_L$ with $L\gg L^\prime$. 
Thus, we can safely employ the following second rule 
when we perform  the averaging of 
an arbitrary function $F(\psm(\bm{X},t),t)$ of the 
small-scale fluctuation  
field $\psm$ multiplied by $\pla$ over the volume $\mal{V}_{L^\prime}$:
\begin{equation}
\mbox{(2)}\hspace{3em} \frac{1}{\mal{V}_{L^\prime}}
\int_{{\cal D}_{L'(X)}}\hspace{-1.5em}d^3X^\prime
\pla_i(\bm{X}^\prime,t;L<\lambda)F(\psm(\bm{X}^\prime,t),t)
=\pla_i(\bm{X},t;L<\lambda) 
\frac{1}{\mal{V}_{L'}}\int_{{\cal D}_{L'}(X)}\hspace{-1.5em}
d^3X^\prime F(\psm(\bm{X}',t),t). \label{eqn:rule2}
\end{equation}
This rule is correct at the lowest order in a  
Taylor expansion of $\pla$:
\begin{equation}
\pla_{i}(\bm{X}',t)=\pla_i(\bm{X},t)+\pla_{i|j}(\bm{X},t)\left(
X'-X\right)_j+\cdots,
\end{equation}
because we have neglected the second term in the above 
compared with the first term. 
In the averaging volume $\mal{V}_{L'}$, the first and 
second terms are of order  $\pla$ and $(\pla L')/L$,
respectively, so the assumption $L^\prime\ll L$ allows us to 
ignore the second term compared with the first term. 
Finally,  we introduce the third 
rule for an arbitrary vector $G_i(\psm)$ which consists of  the 
small-scale displacement 
vector $\psm_i$: 
\begin{eqnarray}
\mbox{(3)}\hspace{3em}\frac{1}{\mal{V}_{L'}}
\int_{{\cal D}_{L'}(X)}\hspace{-1.5em}d^3X^\prime
G_{i|i}(\psm):=\frac{1}{\mal{V}_{L'}}
\int_{\partial \!{\cal D}_{L'}(X)}\hspace{-1.5em}d^3S_i^\prime \ G_i\!\left(\psm\right)
=0\label{eqn:rule3}.
\end{eqnarray}
Here, we have assumed that the $\psm$ are mainly generated  
by random initial density fluctuations with 
only the characteristic wavelength much smaller than $L'$, 
and obeys the periodic boundary condition on the volume 
$\mal{V}_{L'}(\ll \mal{V}_L)$:
\begin{equation}
\kaco{\psm_{i|j}}_{L'}=\frac{1}{\mal{V}_{L'}}
\int_{{\cal D}_{L'}(X)}\hspace{-1.5em}d^3X^\prime
\psm_{i|j}(\bm{X}',t)=0.
\end{equation}
This third rule causes a possible error because we 
neglect the fluctuations with scale $L'\lesssim \lambda \lesssim L$.  
Neglecting the fluctuations with the length comparable with the scale 
$L$ may not be a serious problem, because we have assumed that 
the fluctuations with the scale  $L$ are still in the 
linear regime at the present time. 
However,  neglecting  the fluctuations with scales comparable with $L'$ 
might cause a serious problem. 
For the present we leave this problem open, and we 
consider the situation under the above three rules.

According to these rules, by averaging  both 
sides of Eq.(\ref{eqn:fullevoleq}) over the domain 
$\mal{\cal D}_{L'}$ we obtain
\begin{equation}
3a^2\ddot{a}(t)+a^2\ddot{p}^>_{i|i}(\bm{X},t)+2a\ddot{a}\pla_{i|i}  
+\pla_{i|j}(\bm{X},t)\kaco{\psm_{j|i}\ddot{p}^<_{k|k}-\ddot{p}^<_{j|i} 
\psm_{k|k}}_{L'}\!\!(\bm{X},t)=-4\pi G 
\langle\mal{\rho}\rangle_{L'}(\bm{X};
 L'\lesssim \lambda),\label{eqn:lsdisevol}
\end{equation}
where we have used the following calculation
\begin{eqnarray}
\kaco{2\psm_{j|k}\ddot{p}^<_{k|i}-\psm_{j|i}\ddot{p}^<_{k|k}
-\ddot{p}^<_{j|i}\psm_{k|k}}_{L'}&=&\kaco{2\left(\psm_j\ddot{p}^<
_{k|i}\right)_{|k}-2\psm_j\ddot{p}^<_{k|ik}-\psm_{j|i}\ddot{p}^<_{k|k}
-\ddot{p}^<_{j|i}\psm_{k|k}}_{L'} \nonumber \\
&=&\kaco{-2\left(\psm_j\ddot{p}^<_{k|k}\right)_{|i}+2\psm_{j|i}
\ddot{p}^<_{k|k}-\psm_{j|i}\ddot{p}^<_{k|k}
-\ddot{p}^<_{j|i}\psm_{k|k}}_{L'} \nonumber \\
&=&\kaco{\psm_{j|i}
\ddot{p}^<_{k|k}-\ddot{p}^<_{j|i}\psm_{k|k}}_{L'}\!\!(\bm{X},t).
\end{eqnarray}
Using the Friedmann equation (\ref{eqn:lagfriedac}), 
Eq.(\ref{eqn:lsdisevol})
becomes
\begin{equation}
a^2\ddot{p}^>_{i|i}(\bm{X},t)+2a\ddot{a}\pla_{i|i}(\bm{X},t)  
+\pla_{i|j}(\bm{X},t)\kaco{\psm_{j|i}\ddot{p}^<_{k|k}-\ddot{p}^<_{j|i} 
\psm_{k|k}}_{L'}\!\!(\bm{X},t)=-4\pi G \left(
\langle\mal{\rho}\rangle_{L'}(\bm{X};L'\lesssim\lambda)
-\mal{\rho}_b\right).\label{eqn:zeleqn}
\end{equation}
We emphasize that the source function on the right-hand-side of the 
above equation became the density fluctuation field with a wavelength 
 larger 
than the averaging scale $L'$ and smaller than $L_H$ because of
using the smoothing method. 
Furthermore, the third term on the left-hand-side of 
Eq.(\ref{eqn:zeleqn}) represents a backreaction effect that 
the small-scale non-linear displacement vector $\psm$ has 
on the large-scale perturbation $\pla$. 
Namely, even if the small non-linear displacement vector $\psm_i$
obeys the periodic boundary condition on the volume $\mal{V}_{L'}$, 
the non-linear structures have a possibility to 
give such a backreaction effect on the evolution of 
fluctuations with larger scales in the linear regime of 
the larger fluctuations.

However, we have the following situations where the large-scale 
backreaction becomes zero or negligible compared 
with the other terms.
\begin{itemize}
\item(1) The first case is that 
the  small-scale displacement vector $\psm_i$ can be divided 
into a time function and a spatial function with respect 
to the Lagrangian coordinates: $\psm_i=D(t)\psi^<_i$. 
Then the backreaction term becomes zero:
\[\kaco{\psm_{j|i}\ddot{p}^<_{k|k}-\ddot{p}^<_{j|i} 
\psm_{k|k}}_{L'}=\ddot{D}(t)\kaco{\psi^<_{j|i}\psi^<_{k|k}-
\psi^<_{j|i}\psi^<_{k|k}}_{L'}=0.
\]
During the early stage after the decoupling time of matter and 
radiation, we expect that the $p^<_i$ can be described by the  
Zel'dovich type  solution and thus we may have the above case. 
Actually, Ehlers \& Buchert\cite{ehlersbuchert} have shown that the displacement 
vector at every order has such a separable solution 
 in the Lagrangian perturbation theory.  
But in the non-linear 
regime later around the peak patch of density fluctuation field, 
there is no guarantee that $p^<_i$ is separable.
In this paper, we 
are interested in the non-linear situation of $\psm_i$.
\item(2) The second case is as follows. 
 We can divide  the deformation field of the 
large-scale displacement vector $\pla_{i|j}$
into the divergence, trace-free symmetric, and antisymmetric
 parts without 
loss of generality:
\begin{equation}
\pla_{i|j}=\frac{1}{3}\delta_{ij}\pla_{k|k}+\left(\frac{1}{2}(
\pla_{i|j}+\pla_{j|i})-\frac{1}{3}\delta_{ij}\pla_{k|k}\right)
+\frac{1}{2}(\pla_{i|j}-\pla_{j|i})\equiv
{\cal A} \delta_{ij}+{\cal S}_{ij}+{\cal R}_{ij},
\end{equation}
where
\begin{eqnarray}
&&{\cal A}(\bm{X},t)\equiv \frac{1}{3}\pla_{i|i}, \\
&&{\cal S}_{ij}(\bm{X},t)\equiv \frac{1}{2}(\pla_{i|j}+\pla_{j|i}) 
-\frac{1}{3}\delta_{ij}\pla_{k|k}, \\
&&{\cal R}_{ij}(\bm{X},t)\equiv \frac{1}{2}(\pla_{i|j}-\pla_{j|i}).\label{eqn:Omega}
\end{eqnarray}
As discussed in appendix A, 
we can safely ignore 
the antisymmetric deformation  field 
${\cal R}_{ij}$ compared with the other
     quantities ${\cal A}$ and ${\cal S}_{ij}$ under an appropriate
     assumption: 
${\cal R}_{ij}\approx 0$ or ${\cal R}_{ij}\ll {\cal A},{\cal S}_{ij}$.
Then, if the expansion deformation ${\cal A} $ is assumed to be 
much larger than 
the shear part ${\cal S}_{ij}$, the backreaction term
becomes 
\begin{equation}
\pla_{i|j}\kaco{\psm_{j|i}\ddot{p}^<_{k|k}-\ddot{p}^<_{j|i} 
\psm_{k|k}}_{L'}={\cal A}\delta_{ij}
\kaco{\psm_{j|i}\ddot{p}^<_{k|k}-\ddot{p}^<_{j|i} 
\psm_{k|k}}_{L'}+{\cal S}_{ij}
\kaco{\psm_{j|i}\ddot{p}^<_{k|k}-\ddot{p}^<_{j|i} 
\psm_{k|k}}_{L'}
=O\left(\kaco{{\cal S} \psm\psm}\right).
\end{equation}
Hence, to ignore the backreaction term compared with the other 
terms in Eq.(\ref{eqn:zeleqn}), we have to employ the following 
condition
\begin{equation}
a^2\ddot{{\cal A}}, a\ddot{a}{\cal A} \gg {\cal S}_{ij}
\kaco{\psm_{j|i}\ddot{p}^<_{k|k}-\ddot{p}^<_{j|i} 
\psm_{k|k}}_{L'}\label{eqn:secondcond}.
\end{equation}
This assumption must be handled with  caution.  
Since  $\pla$ represents 
the averaged large-scale field of the original displacement vector
$P_i$, we expect that it can 
 be expressed in the form of Zel'dovich type solution as discussed below. 
Thus, the assumption ${\cal A}\gg {\cal S}_{ij}$ may not be  
 appropriate in the quasi non-linear regime, and we need more
     investigations in detail. 
The condition (\ref{eqn:secondcond}) could be checked in a realistic 
structure formation scenario by using the numerical simulation.
\item(3)The third case is that 
 a locally one-dimensional  motion or a spherical top-hat motion 
dominates for $\psm_{i|j}$ 
in the nonlinear structure formation  on small scales:
\begin{equation}
\psm_{i|j}\approx \psm_{1|1}\delta_{i1}\delta_{j1},
\hspace{2em}\mbox{or}\hspace{2em} \psm_{i|j}
\approx \psm_{k|k}\delta_{ij}.
\end{equation}  
In these  cases, the backreaction term becomes
\begin{eqnarray}
\pla_{i|j}\kaco{\psm_{j|i}\ddot{p}^<_{k|k}-\ddot{p}^<_{j|i} 
\psm_{k|k}}_{L'}&\approx& \sum_{\mbox{clump}}\pla_{i|j}
\kaco{\psm_{j|i}\ddot{p}^<_{k|k}-\ddot{p}^<_{j|i} 
\psm_{k|k}}_{L'}\nonumber \\
&=&\sum_{\mbox{clump}}\pla_{1|1}
\kaco{\psm_{1|1|}\ddot{p}^<_{1|1}-\ddot{p}^<_{1|1} 
\psm_{1|1}}_{L'}\hspace{1.5em}\mbox{or}\hspace{1.5em}\sum_{\mbox{clump}}
\pla_{l|l}\delta_{ij}\kaco{\psm_{j|i}\ddot{p}^<_{k|k}
-\ddot{p}^<_{j|i}\psm_{k|k}}_{L'} \nonumber \\
&=&0,
\end{eqnarray}
where $\sum_{\mbox{clump}}$ denotes  
the sum of the number of 
 non-linear small-scale clumps included into 
the averaging domain $\mal{\cal D}_{L'}$. 
Here we have assumed that we can  replace the integral in  $\kaco{\cdots}_{L'}$ 
with the sum over the clumps.  

\item(4) Finally, we remark that the backreaction term can be 
rewritten as 
\begin{equation}
\pla_{i|j}\kaco{\psm_{j|i}\ddot{p}^<_{k|k}-\ddot{p}^<_{j|i} 
\psm_{k|k}}_{L'}=\pla_{i|j}\frac{d}{dt}\kaco{\psm_{j|i}\dot{p}^<_{k|k}
-\dot{p}^<_{j|i}\psm_{k|k}}_{L'}. \label{eqn:rewrittenback} 
\end{equation}
This suggests that we may employ a time averaging together with the
spatial average to obtain the evolution equation of the large-scale
dynamics. Once the nonlinear structures are developed, its 
time scale will be much shorter than that of large-scale dynamics. 
Thus, the averaging over the local time scale will eliminate 
 the above backreaction term. Although this is interesting, we
will not pursue it here and leave it for future study. 
\end{itemize}

Under these situations or some combined situation of them, 
the evolution equation (\ref{eqn:zeleqn}) 
for the large-scale displacement vector $\pla_i$ yields
\begin{equation}
a^2\ddot{p}^>_{i|i}+2a\ddot{a}\pla_{i|i}+O\left((\pla)^2\right)
=-4\pi G \mal{\rho}_b
\mal{\delta}_{L'}\!\!(\bm{X};L'\lesssim\lambda),\label{eqn:usualzel}
\end{equation}
where 
\begin{equation}
\mal{\delta}_{L'}\!\!(\bm{X};L'\lesssim\lambda)\equiv
\frac{\langle\mal{\rho}\rangle_{L'}(\bm{X};L'\lesssim\lambda)
-\mal{\rho}_b}{\mal{\rho}_b}.\label{eqn:defmediumdens}
\end{equation}
We again note that, since $\pla_i$ is defined by smoothing 
the original displacement vector $P_i$, we need only 
the lowest order of $\pla$.

Eq.(\ref{eqn:usualzel}) entirely agrees with the first order 
evolution equation in the Lagrangian perturbation theory, 
so we can proceed to solve it for $\pla_i$ 
iteratively using the 
solution of the scale factor $a(t)$ for the Friedmann background 
equation (\ref{eqn:lagfriedac}). In this way, introducing 
the artificial cutoff length $L$ and the spatial averaging over 
the volume $\mal{V}_{L'}$ smaller than the horizon scale allow 
us to understand the validity of  the 
Lagrangian perturbation theory  in
investigating 
the large-scale structure formation even if the universe has non-linear 
structures on small scales.  
Our formalism clarifies  the meaning of the 
optimized or truncated Lagrangian perturbation theory which has  
been frequently used. Many authors\cite{Coles,hamana,Melott94,Melott95,weiss} 
have shown that the optimized or truncated 
Lagrangian perturbation theory reproduces the results of the
large-scale structure larger than the smoothing scale 
by the full N-body simulation.  
This indicates that some of 
the above assumptions may be satisfied 
 in the situation. 
It may be interesting to investigate which assumption is valid, and 
this will be presented elsewhere.

\subsection{The local small-scale non-linear evolution equations 
in the Lagrangian picture}

Next we derive the local evolution equation
for the small-scale displacement vector $\psm_i$. 
We have obtained the background equation and 
the evolution equation for the large-scale displacement 
vector within the Lagrangian framework. We want to construct 
the local evolution equation for the small-scale displacement 
vector including the effect of the gravitational 
instability of the surrounding large-scale structure.  
If we do so, we may have the possibility to use the high-resolution 
N-body simulation or the semi-analytic approaches 
 only for the small-scale non-linear 
structure formation and to use Zel'dovich's approximation 
for the large-scale 
structure formation, simultaneously. 

Subtracting Eq.(\ref{eqn:zeleqn}) from Eq.(\ref{eqn:fullevoleq}), 
we obtain the following local small-scale evolution equation in the
situations taken up in the previous subsection:
\begin{eqnarray}
&&a^2\ddot{p}^<_{i|i}+2a\ddot{a}\psm_{i|i}+
a\left(p^>_{i|i}\ddot{p}^<_{j|j}-p^>_{i|j}\ddot{p}^<_{j|i}
\right)+a\left(\ddot{p}^>_{i|i}p^<_{j|j}-\ddot{p}^>_{i|j}
p^<_{j|i}\right)+
a\left(\psm_{i|i}\ddot{p}^<_{j|j}-\psm_{i|j}\ddot{p}^<_{j|i}
\right)+\frac{\ddot{a}}{2}\left(\psm_{i|i}\psm_{j|j}
-\psm_{i|j}\psm_{j|i}\right)\nonumber \\
&&\hspace{7em}+\ddot{a}\left(p^>_{i|i}p^<_{j|j}-p^>_{i|j}p^<_{j|i}\right)
+p^>_{i|i}\left(p^<_{j|j}\ddot{p}^<_{k|k}
-p^<_{k|j}\ddot{p}^<_{j|k}\right)
+p^>_{i|j}\left(2p^<_{j|k}\ddot{p}^<_{k|i}-p^<_{k|k}\ddot{p}^<_{j|i}
-\ddot{p}^<_{k|k}p^<_{j|i}\right)\nonumber \\
&&\hspace{7em}+\frac{1}{2}\ddot{p}^>_{i|i}\left(p^<_{j|j}p^<_{k|k}
-p^<_{k|j}p^<_{j|k}\right)
+\ddot{p}^>_{i|j}\left(\psm_{j|k}\psm_{k|i}
-\psm_{k|k}\psm_{j|i}\right)\nonumber \\
&&\hspace{14em}+\frac{1}{2}\epsilon_{ijk}\epsilon_{abc}\psm_{i|a}
\psm_{j|b}\ddot{p}^<_{k|c}=
-4\pi G
\langle\mal{\rho}\rangle_{L'}\mal{\delta}\!(\bm{X};\lambda\lesssim
L'),
\label{eqn:localeveq}
\end{eqnarray}
where
\begin{equation}
\mal{\delta}(\bm{X};\lambda\lesssim L')\equiv \frac{
\mal{\rho}\!(\bm{X};\lambda)-\langle\mal{\rho}\rangle_{L'}(\bm{X};
L'\lesssim\lambda)}{\langle\mal{\rho}\rangle_{L'}(\bm{X};
L'\lesssim\lambda)}.
\end{equation}
Note that the source function is a density fluctuation 
field with scale smaller than $L'$.

If we could solve Eq.(\ref{eqn:usualzel}) for the 
$\pla_i$, we could  solve Eq.(\ref{eqn:localeveq}) for
the local displacement vector $\psm_i$ in principle 
by substituting  the solution $\pla$. 
However, in practice, it will be very difficult to do so because 
Eq.(\ref{eqn:localeveq}) is a highly nonlinear differential equation. 
Instead we restrict ourself to a  more simple situation in this paper 
in order to see clearly the environmental effect on the small-scale 
dynamics. 
Namely, we consider the second situation described in the previous 
subsection. In this case the Lagrangian divergence of 
the displacement field  dominates on the averaged large-scale
dynamics. It is straightforward to extend the formulation 
 taking into account 
the effect of tracefree part of the large-scale deformation field which 
represents the surrounding tidal field, 
and it will be  presented elsewhere. Here, we consider only  the simplest case.  
Then we  can rewrite Eq.(\ref{eqn:localeveq}) as 
\begin{eqnarray}
&&\left(a+{\cal A}\right)^2\ddot{p}^<_{i|i}+2(a+{\cal A})(\ddot{a}+\ddot{{\cal A}})
\psm_{i|i}+(a+{\cal A})\left(\psm_{i|i}\ddot{p}^<_{j|j}-\psm_{i|j}
\ddot{p}^<_{j|i}\right)\nonumber \\
&&\hspace{8em}+\frac{1}{2}(\ddot{a}+\ddot{{\cal A}})\left(\psm_{i|i}
\psm_{j|j}-\psm_{i|j}\psm_{j|i}\right)+
\frac{1}{2}\epsilon_{ijk}\epsilon_{abc}\psm_{i|a}
\psm_{j|b}\ddot{p}^<_{k|c}=-4\pi G
\langle\mal{\rho}\rangle_{L'}\mal{\delta}\!\!(\bm{X};\lambda\lesssim L').
\end{eqnarray}
Or, using Eqs.(\ref{eqn:averfriedmannacce}), (\ref{eqn:usualzel}) and 
(\ref{eqn:defmediumdens}), this equation becomes
\begin{eqnarray}
&&3\left(a+{\cal A}\right)^2\left(\ddot{a}+\ddot{{\cal A}}\right)
+\left(a+{\cal A}\right)^2\ddot{p}^<_{i|i}+2(a+{\cal A})(\ddot{a}+\ddot{{\cal A}})
\psm_{i|i}+(a+{\cal A})\left(\psm_{i|i}\ddot{p}^<_{j|j}-\psm_{i|j}
\ddot{p}^<_{j|i}\right)\nonumber \\
&&\hspace{8em}+\frac{1}{2}(\ddot{a}+\ddot{{\cal A}})\left(\psm_{i|i}
\psm_{j|j}-\psm_{i|j}\psm_{j|i}\right)+
\frac{1}{2}\epsilon_{ijk}\epsilon_{abc}\psm_{i|a}
\psm_{j|b}\ddot{p}^<_{k|c}=-4\pi G \mal{\rho}\!(\bm{X};\lambda),
\end{eqnarray}
Note that we have normalized the global scale factor to  $\mal{a}=1$ 
in this section and considered the Einstein-de Sitter background. 
If the above equation is compared with 
the original equation (\ref{eqn:originaleq2}) in the 
Lagrangian picture, it is seen that  
the effect of the large-scale structure formation 
on the small non-linear scales is represented 
as a modification of the global scale factor 
$(a+ {\cal A})(t)$ on the surrounding large scale.

Thus, since the spatial gradient of 
the expansion ${\cal A}$ with respect to 
the Lagrangian coordinates can be ignored inside the 
volume $\mal{V}_{L'}$ in this  case,                        
the above equation corresponds to the following 
system of Eulerian equations on the comoving coordinates $y=x/(a+{\cal A})$:
{
\setcounter{enumi}{\value{equation}}
\addtocounter{enumi}{1}
\setcounter{equation}{0}
\renewcommand{\theequation}{\thesection.\theenumi\alph{equation}}
\begin{eqnarray}\label{eqn:effefried}
&&\frac{\partial \rho}{\partial t}+3\frac{\dot{a}+\dot{{\cal A}}}
{a+{\cal A}}\rho+\frac{\partial }{\partial y^i}(\rho u^i)=0,\\
&&\frac{\partial u^i}{\partial t}+2\frac{\dot{a}+\dot{{\cal A}}}
{a+{\cal A}}u^i+u^j\frac{\partial u^i}{\partial y^j}=
-\frac{1}{(a+{\cal A})^2}\frac{\partial \phi}{\partial y^i},\\
&&\Delta_y\phi=4\pi G(a+{\cal A})^2\left(\rho(\bm{y},t)
-\kaco{\rho}_{L'}\!\!(t)\right),
\end{eqnarray}\setcounter{equation}{\value{enumi}}}\noindent
where $u^i$ denotes the peculiar velocity on the 
comoving coordinates $\{\bm{y}\}$ and the physical peculiar velocity 
is $(a+{\cal A})u^i$, and $\kaco{\rho}_{L'}\!\!(t)$ represents the effective 
background density defined by $\kaco{\rho}_{L'}\!\!(t)\equiv
\langle\mal{\rho}\rangle_{L'}/(a+{\cal A})^3$.  
The comoving coordinates $\{\bm{y}\}$ should be defined on 
the Eulerian coordinates  at the initial time, when 
fluctuations with all scales are much smaller than unity. 
Thus, we can interpret the system of equations 
(\ref{eqn:effefried}-c) as describing the evolutions of small-scale 
fluctuations on the {\it effective} FRW background model  
 characterized by the modified scale factor $(a(t)+{\cal A}(t))$. 
This equation  means that, because of 
the fact ${\cal A}<0$ within  the large-scale
objects that  are collapsing, the fluctuations on smaller 
scales tend to collapse earlier  than 
the fluctuations in the large-scale void where one  has ${\cal A}>0$. 
Furthermore, we can perform the N-body simulation 
using the above Eulerian set of equations inside 
the box with volume $\mal{V}_{L'}$ 
in the  usual way only if  the scale factor is 
modified to the effective scale factor $(a+{\cal A})$ inside the 
box by using the Zel'dovich solution or the improved version of it 
for ${\cal A}(t)$. 
In this way, our formalism will be useful for the consideration of 
the environmental  effects on the behavior of the 
fluctuations.

\section{Discussion}

  We have developed a formalism which allows us to investigate the
relation between large-scale quasi-linear dynamics and small-scale
nonlinear dynamics  using the averaging method in the Lagrangian
theory. 
We have derived the coupled equation for the large-scale dynamics and 
the small-scale dynamics.   
In the case where the averaged large-scale dynamics is expansion
dominated, we have shown that the large-scale dynamics decouples 
from  the small-scale nonlinear dynamics. Then, on the other hand,
the small-scale dynamics is influenced by the large-scale dynamics 
in such a way  that  the local small-scale equations contain 
the modified scale factor 
of the large scale. The modified scale factor is the sum of the global
scale factor and the expansion of the region considered. 
Our result strongly suggests that there will be more 
complicated environmental  effects in
local small-scale dynamics which one cannot ignore.
 
There may be several possibilities to generalize our analysis.  
One would be to employ some approximation to solve the local dynamics 
in the Lagrangian framework. For example the spherical symmetric
approximation on small scales may be reasonable and we may employ 
the Press-Schechter\cite{press} type approach to the local region\cite{lee,MW,monaco}. 
More challenging would be 
to employ N-body simulation on small scales. We have mentioned this
possibility in the previous section, but it seems more work is
necessary to achieve this consistently within our scheme.
One thing missing in our formalism is the mutual gravitational
interaction between the small clumps contained in a large scale
environment.
This is because we have ignored the dynamical freedom between 
$L'<\lambda<L$. 
Also, although we have here 
ignored the large-scale tidal effect on the small-scale
dynamics in Eq.(\ref{eqn:effefried}-c) for the purpose of a simple illustration, 
it may play an important role in the 
hierarchical structure formation. For example,  
Bond \& Myers\cite{bond} have
investigated 
the important influence  of the tidal effect for the merging history of 
halo objects. It is straightforward to extend our formalism
including the large-scale tidal effect on the small-scale 
dynamics, and 
investigating the environmental influence will be interesting. 
We would like to come back these points in future study.

\section*{Acknowledgment}
Part of this study is done during our stay at Max-Planck Institute for
Astrophysics whose hospitality is acknowledged. We would like to thank
Gerhard B\"orner and Thomas Buchert for useful discussions and
valuable suggestions.

\begin{appendix}
\section{The large-scale vorticity  mode in the Lagrangian
picture}

 In this appendix, we investigate  the behavior of the 
rotational deformation field  of the large-scale displacement 
vector $\pla_i$. 
For this purpose, it is convenient to make use of 
the {\it Kelvin's circulation transport equation}
(\ref{eqn:intomega}) in the Lagrangian picture:
\begin{equation}
\frac{1}{2}\epsilon_{blc}f_{j|b}\dot{f}_{j|c}=\mal{\omega}_l.
\end{equation}
By substituting the full trajectory field $f_i=aX_i+
\pla_i+\psm_i$ into the above equation, 
we obtain
\begin{equation}
\frac{1}{2}\epsilon_{ijk}\left[a(\dot{p}^>_{j|k}+
\dot{p}^<_{j|k})-\dot{a}(\pla_{j|k}+\psm_{j|k})+
\pla_{l|j}\dot{p}^<_{l|k}
+\dot{p}^>_{l|k}\psm_{l|j}+\psm_{l|j}\dot{p}^<_{l|k}
\right]=-\mal{\omega}_i(\bm{X};l<\lambda<L_H).
\end{equation}
If we perform the averaging of the above equation 
over the domain $\mal{{\cal D}}_{L'}$ according to the rules 
(\ref{eqn:rule1}), (\ref{eqn:rule2}) and (\ref{eqn:rule3}), 
we  get 
\begin{equation}
\frac{1}{2}\epsilon_{ijk}\frac{d}{dt}\left(\frac{\pla_{j|k}}
{a}\right)=-\frac{1}{a^2}\langle\mal{\omega}_i\rangle_{L'}(\bm{X};L'\lesssim\lambda<L_H).
\label{eqn:appomega}
\end{equation}
In the above derivation, for example, we have used the calculation such
as 
\begin{eqnarray}
\frac{1}{2}\epsilon_{ijk}\kaco{\pla_{l|j}\dot{p}^<_{l|k}}_{L'}=
\frac{1}{2}\epsilon_{ijk}\kaco{(\pla_{l|j}\dot{p}^<_l)_{|k}}_{L'}=0.
\end{eqnarray}
By noting $\pla_{i|j}(\bm{X},t_I)=0$ and 
using the variable ${\cal R}_{ij}$ defined by Eq.(\ref{eqn:Omega}), 
Eq.(\ref{eqn:appomega}) can be integrated as
\begin{equation}
\frac{1}{2}\epsilon_{ijk}{\cal R}_{jk}=
-a\langle\mal{\omega}_i\rangle_{L'}\int^t_{t_I}dt\frac{1}{a^2\!(t^\prime)}.
\end{equation}
Thus, if we assume that the averaged initial 
 large-scale 
vorticity field $\kaco{\omega_i}_{L'}$ is exactly  zero, 
we arrive at the conclusion
\begin{equation}
{\cal R}_{ij}=0.
\end{equation}  
According to  this consideration, even if we do not adopt the above  assumption,  
we can safely ignore 
the large-scale rotational field ${\cal R}_{ij}$ compared 
with the 
trace part of the 
 deformation field  ${\cal A}$ and the trace-free symmetric part  ${\cal S}_{ij}$,
because the initial vorticity field is much smaller than the expansion 
field and the shear field based on the linearized theory\cite{peebles}.
\end{appendix}

\end{document}